\documentstyle[12pt,epsfig]{article}
\makeatletter
\def\slash#1{{\mathpalette\c@ncel{#1}}} 
\makeatother

\newcommand\beq{\begin{eqnarray}}
\newcommand\eeq{\end{eqnarray}}

\newcommand\la{\langle}
\newcommand\ra{\rangle}

\def\pslash{\rlap/{\mkern-1mu p}}

\def\kslash{\slash{\mkern-1mu k}}

\def\xhat{\widehat{x}}
\def\zhat{\widehat{z}}

\begin{document}
\begin{flushright}
\end{flushright}
\vspace*{15mm}
\begin{center}
{\Large \bf Master Formula for the Three-Gluon
Contribution to Single Spin Asymmetry 
\\[2mm] 
in Semi-Inclusive Deep Inelastic Scattering}
\vspace{1.5cm}\\
 {\sc Yuji Koike$^1$, Kazuhiro Tanaka$^2$ and Shinsuke Yoshida$^3$}
\\[0.4cm]
\vspace*{0.1cm}{\it $^1$ Department of Physics, Niigata University,
Ikarashi, Niigata 950-2181, Japan}\\
\vspace*{0.1cm}{\it $^2$ Department of Physics, 
Juntendo University, Inzai, Chiba 270-1695, Japan}\\
\vspace*{0.1cm}{\it $^3$ Graduate School of Science and Technology, Niigata University,
Ikarashi, Niigata 950-2181, Japan}
\\[3cm]

{\large \bf Abstract} \end{center}

We derive a ``master formula" for the contribution of the three-gluon
correlation function in the nucleon to the twist-3 single-spin-dependent cross
section for semi-inclusive deep-inelastic scattering, $ep^\uparrow\to eDX$.   
This is an extension of the similar formula known for the
so-called soft-gluon-pole contribution induced by the quark-gluon correlation function
in a variety of processes.
Our master formula 
reduces 
the relevant interfering partonic subprocess with the participation of the three gluons to 
the Born cross sections for the $\gamma^*g\to c\bar{c}$ scattering, 
which reveals the new structure behind the twist-3 single spin asymmetry and simplifies
the actual calculation greatly.   A possible extension to higher order corrections
is also discussed.

\newpage
\section{Introduction}

Single (transverse) spin asymmetries (SSA) in high-energy inclusive processes, such as
semi-inclusive deep inelastic scattering (SIDIS), Drell-Yan process, and hadron production
in $pp$ collision, 
appear as a consequence of multi-parton correlations
in the hadrons, which did not show up in the naive parton picture for high-energy scattering.  
SSA reveals the multi-parton correlation most unambiguously as a leading twist-3 observable
in the framework of collinear factorization which is valid to describe
hadron production with large transverse momentum $P_T\gg \Lambda_{\rm QCD}$.  
\footnote{
For the contribution from the quark-gluon correlation functions, 
consistency between this approach and the transverse-momentum-dependent factorization 
in the region $\Lambda_{\rm QCD}\ll P_T \ll Q$ with the hard scale $Q$
has also been shown for some processes\,\cite{JQVY06,JQVY06DIS,KVY08}. }
Among those correlations, the effect of quark-gluon correlations
have been widely studied in the literature and our understanding on the mechanism of SSA has made
a great progress\,\cite{ET82}-\cite{Koike:2009yb}.   
There are also some studies on the contribution to SSA from the
purely gluonic correlation in the nucleon\,\cite{KQ08,KQVY08,BKTY10,Koike:2010jz} 
and the multi-parton correlations 
in the fragmentation functions\,\cite{YZ09,Kang:2010zzb}. 

In our recent paper\,\cite{BKTY10}, we have established the formalism 
for calculating the
twist-3 single-spin dependent cross section induced by the
three-gluon correlation functions in the transversely polarized nucleon.  
There, we clarified a complete set of the gauge-invariant three-gluon correlation functions
and derived the corresponding cross section for the $D$-meson production
in SIDIS, $ep^\uparrow\to eDX$,
which is relevant to probe the purely gluonic correlation leading to SSA.
\footnote{The $D$-meson production at large $P_T$ in SIDIS is dominated by
the photon-gluon fusion subprosess.
Other possible sources of the asymmetry in the $D$-meson production are associated with
the charm quark content in the polarized nucleon,
such as the $c$-quark transversity distribution and the
$c$-quark-gluon correlation functions, but those intrinsic charm contributions
are expected to give tiny corrections in the large $P_T$ region which we are interested in this paper.}
In particular, we have proved the factorization property of the twist-3 cross section
and given a detailed prescription of expressing the cross section in terms of the
gauge-invariant three-gluon correlation functions.  
Our result differed from the previous study~\cite{KQ08,KQVY08}, and 
we clarified why the previous result needs to be corrected.  

The partonic cross section in this formalism is given as a pole contribution
of an internal propagator in the hard part, reflecting the naively $T$-odd nature 
of SSA.  The pole forces one of the gluon lines
in the three-gluon correlation functions to be soft, 
and hence the pole is called the soft-gluon-pole (SGP).
For the SGP contribution from the quark-gluon correlation functions,
it has been shown in \cite{KT071,KT072} that
the corresponding twist-3 hard cross sections for $ep^\uparrow\to e\pi X$ and $p^\uparrow p\to \pi X$
have a simple relation with the twist-2 unpolarized hard cross section
for the same processes.
This connection greatly facilitates the actual calculation
and makes the structure of the SGP contribution transparent.

The purpose of this paper is to show that the whole three-gluon contribution
to $ep^\uparrow \to eDX$
can also be obtained from 
the Born cross sections for the 
$\gamma^*g\to c\bar{c}$ scattering.  
In particular, we will show that some of the twist-3 cross section is completely determined
by the gluon-contribution to the twist-2 unpolarized cross section.

The remainder of this paper is organized as follows:
In section 2, we briefly recall a complete set of the
twist-3 three-gluon correlation functions in the transversely polarized nucleon.  
In section 3, we summarize the twist-3 formalism for 
the three-gluon correlation functions developed in \cite{BKTY10}.  
We mostly follow the notation used in \cite{BKTY10}, but introduce a
slightly different convention for the azimuthal angles and the hard part, 
which turned out to be more convenient for our purpose.  
In section 4, we develop a new master formula which connects the three-gluon contribution
to the twist-3 single-spin-dependent cross section to
a simpler cross section for the $\gamma^*g\to c\bar{c}$ scattering.
We also discuss the extension of the formula for higher order corrections.  
Section 5 is devoted to a brief summary of the outcome of this paper.

\section{Three-gluon correlation functions in the transversely-polarized nucleon}

As clarified in \cite{BJLO,Braun09,BKTY10}, there are two-independent
twist-3 three-gluon correlation functions in the transversely-polarized nucleon, 
$O(x_1,x_2)$ and $N(x_1,x_2)$, 
which are the Lorentz-scalar functions of the two momentum fractions $x_1$ and $x_2$, defined as 
\beq
&&\hspace{-0.8cm}O^{\alpha\beta\gamma}(x_1,x_2)
=-gi^3\int{d\lambda\over 2\pi}\int{d\mu\over 2\pi}e^{i\lambda x_1}
e^{i\mu(x_2-x_1)}\la pS|d^{bca}F_b^{\beta n}(0)F_c^{\gamma n}(\mu n)F_a^{\alpha n}(\lambda n)
|pS\ra 
\nonumber\\
&&=2iM_N\left[
O(x_1,x_2)g^{\alpha\beta}\epsilon^{\gamma pnS}
+O(x_2,x_2-x_1)g^{\beta\gamma}\epsilon^{\alpha pnS}
+O(x_1,x_1-x_2)g^{\gamma\alpha}\epsilon^{\beta pnS}\right]
\label{3gluonO},\\
&&\hspace{-0.8cm}N^{\alpha\beta\gamma}(x_1,x_2)
=-gi^3\int{d\lambda\over 2\pi}\int{d\mu\over 2\pi}e^{i\lambda x_1}
e^{i\mu(x_2-x_1)} \la pS|if^{bca}F_b^{\beta n}(0)F_c^{\gamma n}(\mu n)F_a^{\alpha n}(\lambda n)
|pS\ra
\nonumber\\
&&=2iM_N\left[
N(x_1,x_2)g^{\alpha\beta}\epsilon^{\gamma pnS}
-N(x_2,x_2-x_1)g^{\beta\gamma}\epsilon^{\alpha pnS}
-N(x_1,x_1-x_2)g^{\gamma\alpha}\epsilon^{\beta pnS}\right], 
\label{3gluonN}
\eeq
up to the irrelevant terms of twist-4 and higher,
where 
$F_a^{\alpha n}\equiv F_a^{\alpha \beta}n_{\beta}$ 
with $F_a^{\alpha\beta}=\partial^\alpha A^\beta_a
-\partial^\beta A^\alpha_a +gf_{abc}A_b^\alpha A_c^\beta$ being the gluon field strength tensor,
$d^{bca}$ and $f^{bca}$ are, respectively, the symmetric
and anti-symmetric structure constants of the color SU(3) group,
and we have suppressed the gauge-link operators which appropriately connect the field strength 
tensors so as to ensure
the gauge invariance.
$p$ is the nucleon momentum, 
$S$ is the transverse spin vector of the
nucleon normalized as $S^2=-1$, and  $M_N$ is the nucleon mass
so that $O(x_1,x_2)$ and $N(x_1,x_2)$ are dimensionless.
In the twist-3 accuracy,
$p$ can be regarded as lightlike ($p^2=0$) and 
$n$ is another lightlike vector satisfying $p\cdot n=1$.  To 
to be specific, we take $p^\mu=(p^+,0, \mathbf{0}_\perp)$, $n^\mu=(0,n^-, \mathbf{0}_\perp)$ and 
$S^\mu =(0,0, \mathbf{S}_\perp)$.
Hermiticity, invariance under the transformations $P$ and $T$, and the permutation symmetry among the participating three
gluon-fields imply that
$O(x_1,x_2)$ and $N(x_1,x_2)$ are real functions and satisfy the relations,
\beq
&&O(x_1,x_2)=O(x_2,x_1),\qquad O(x_1,x_2)=O(-x_1,-x_2),\nonumber\\
&&N(x_1,x_2)=N(x_2,x_1),\qquad N(x_1,x_2)=-N(-x_1,-x_2).\label{symN}  
\eeq

\section{Summary of the twist-3 formalism for $ep^\uparrow\to eDX$}

\subsection{Kinematics}

Here, we summarize the kinematics for the SIDIS process,
\beq
e(\ell)+p^\uparrow(p, S) \to e(\ell')+D(P_h)+X, 
\label{SIDIS}
\eeq
where 
the final $D$-meson has the mass $m_h$, i.e., $P_h^2=m_h^2$. 
This process is described by
the five independent Lorentz invariants: 
\beq
S_{ep}=(p+\ell)^2,\quad
x_{bj}={Q^2\over 2p\cdot q},\quad Q^2=-q^2=-(\ell-\ell')^2,
\quad z_f={p\cdot P_h\over p\cdot q},\quad
q_T=\sqrt{-q_t^2},  
\eeq
where $q_t$ is the ``transverse" component of $q$ 
defined as
\beq
q_t^\mu= q^\mu+\left(\frac{m_h^2p\cdot q}{\left(p\cdot P_h\right)^2} 
- {P_h\cdot q\over p\cdot P_h}\right)p^\mu -{p\cdot q\over p\cdot P_h} P_h^\mu,
\eeq
satisfying 
$q_t\cdot p=q_t\cdot P_h=0$.  
In the actual calculation we work in
the hadron frame
where the virtual photon and the initial nucleon are
collinear, i.e., both move along the $z$-axis. 
In this frame, their momenta $q$ and $p$ are given by
\beq
q^\mu = (q^0, \vec{q})=(0,0,0,-Q),\qquad p^\mu = \left( {Q\over 2x_{bj}},0,0,{Q\over 2x_{bj}}\right) \; . 
\label{pmu}
\eeq
The azimuthal angle of the hadron plane as measured from the $xz$ plane
is taken to be
$\chi$ and thus the momentum of the $D$-meson is parameterized as
\beq
P_h^\mu = {z_f Q \over 2}\left( 1 + {q_T^2\over Q^2}+ {m_h^2\over z_f^2Q^2},{2 q_T\over Q}\cos\chi,
{2 q_T\over Q}\sin\chi,
-1+{q_T^2\over Q^2}+{m_h^2\over z_f^2Q^2}\right) \; .
\label{Dmomem}
\eeq
The transverse momentum of the $D$-meson in this 
frame is given by $P_{hT}=z_f q_T$, which is true in any frame where the
3-momenta $\vec{q}$ and $\vec{p}$ 
are collinear.  The azimuthal angle of the lepton plane measured from the $xz$ plane is taken to be $\phi$ and 
thus the lepton momentum can be parameterized as
\beq
\ell^\mu={Q\over 2}\left( \cosh\psi,\sinh\psi\cos\phi,
\sinh\psi\sin\phi,-1\right) \; ,\nonumber\\
\ell'^\mu={Q\over 2}\left( \cosh\psi,\sinh\psi\cos\phi,
\sinh\psi\sin\phi,1\right) \; ,
\label{eq2.lepton}
\eeq
where
\beq
\cosh\psi = {2x_{bj}S_{ep}\over Q^2} -1 \; .
\label{eq2.cosh}
\eeq
We parameterize the transverse spin vector of the initial nucleon $S^\mu$
as
\beq
S^\mu= (0,\cos\Phi_S,\sin\Phi_S,0),
\label{phis}
\eeq
with the azimuthal angle $\Phi_S$ of $\vec{S}$.  
Although three azimuthal angles $\phi$, $\chi$ and $\Phi_S$ are defined above, 
it is obvious that 
the cross section for $ep^\uparrow\to eD X$ depends on them through only the relative angles 
$\phi-\chi$ and $\Phi_S-\chi$.  
Thus, it can be expressed in terms of
$S_{ep}$, $x_{bj}$, $Q^2$, $z_f$, $q_T^2$, $\phi-\chi$ and $\Phi_S-\chi$ in the above hadron
frame.
Note that $\phi$, $\chi$ and $\Phi_S$ are invariant under boosts in the 
$\vec{q}$-direction, so that the cross section presented below is
the same in any frame where $\vec{q}$ and $\vec{p}$ are collinear.  

With the kinematical variables defined above, 
the differential cross section for 
the $D$-meson production in SIDIS using the unpolarized lepton
can be calculated with
\beq
{d^6 \sigma \over
d x_{bj}dQ^2dz_f dq_T^2 d\phi d\chi}
={\alpha_{em}^2 \over 128\pi^4 x_{bj}^2 S_{ep}^2 Q^2}
z_f L^{\mu\nu}(\ell, \ell')W_{\mu\nu}(p,q,P_h),
\label{diffsigma}
\eeq
where $L^{\mu\nu}(\ell, \ell')=2(\ell^\mu \ell'^\nu + \ell^\nu \ell'^\mu)-Q^2g^{\mu\nu}$
is the corresponding leptonic tensor,
$W_{\mu\nu}(p,q,P_h)$ is the hadronic tensor in the same normalization as in \cite{BKTY10}, and  
$\alpha_{em}=e^2/(4\pi)$ is the QED coupling constant.  
The single-spin ($\vec{S}$) dependent part in the cross section (\ref{diffsigma})
describes the SSA in the process (\ref{SIDIS}).
For this part, one can transform the azimuthal element in the LHS of (\ref{diffsigma})
as $d\phi d\chi\to d\phi d\Phi_S$ and set $\chi=0$.
In fact, in our previous paper \cite{BKTY10}, 
$\phi$ and $\Phi_S$ were used as the azimuthal angles, respectively, for
the lepton plane and the spin vector measured from the hadron plane by setting $\chi=0$ from the beginning.  
In the expression for the differential cross section derived in \cite{BKTY10}, 
the differential element ``$d\Phi_S$" appearing in the above sense was missing, and thus should be
supplied together with the factor $1/(2\pi)$ for the cross section
with all the results unchanged.

\subsection{Twist-3 cross section}

As shown in \cite{BKTY10}, 
the contribution from the three-gluon correlation functions 
to the hadronic tensor $W_{\mu\nu}$ 
is relevant for having the sizeable SSA 
in the collinear factorization to describe the large-$P_{hT}$ $D$-meson production 
and is represented by the diagrams of the type shown in Fig.~1.   
Here,
the twist-2 fragmentation function $D(z)$ for a $c$-quark to become the $D$-meson
is already factorized as the upper blob.
In the other part of the diagrams,
the partonic hard part $S^{abc}_{\mu\nu;\rho\tau\lambda}(k_1,k_2,q,p_c)$, represented by the middle blob,
is combined with 
the correlation functions $\sim \la  A_b^\tau A_c^\lambda A_a^\rho \ra$ of the gluon fields, $A_a^\rho(\xi)$, 
in the nucleon (lower blob), 
where $\rho,\,\tau,\,\lambda$ and $a,\,b,\,c$ are, respectively,
Lorentz and color indices for the relevant three gluon-fields.  
$\mu$ and $\nu$ in $S^{abc}_{\mu\nu;\rho\tau\lambda}(k_1,k_2,q,p_c)$ represent the
Lorentz indices for the virtual photon.  
Compared to 
the conventional
photon-gluon fusion subprocess, $\gamma^*g\to c\bar{c}$,
an additional gluon participates in the partonic hard part $S^{abc}_{\mu\nu;\rho\tau\lambda}(k_1,k_2,q,p_c)$
and allows us to obtain interfering phase arising from 
the unpinched pole contribution of an internal propagator 
contained in $S^{abc}_{\mu\nu;\rho\tau\lambda}(k_1,k_2,q,p_c)$.
Owing to the symmetry property of the spin-dependent part of the 
nucleon matrix elements 
$\la A_b^\tau A_c^\lambda A_a^\rho \ra$
under the $P$- and $T$-transformations~\cite{BKTY10},
only such interfering contribution
in the hard part $S^{abc}_{\mu\nu;\rho\tau\lambda}(k_1,k_2,q,p_c)$ 
can give rise to the single-spin-dependent cross section.
In the leading order with respect to the QCD coupling constant,
the corresponding contribution to
$S^{abc}_{\mu\nu;\rho\tau\lambda}(k_1,k_2,q,p_c)$ can be obtained from the 
diagrams shown in Fig. 2 (together with their mirror diagrams), where 
the bar on a propagator indicates that the pole part
is to be taken from the propagator; in principle, other propagators can produce 
similar pole contributions, but we need not consider separately those contributions, which indeed
correspond to the diagrams obtained by the permutation of the gluons in the diagrams of Fig.~2.
It has been shown in \cite{BKTY10} that
the total contribution to the single-spin-dependent cross section from Figs.~1, 2
can be expressed in a gauge-invariant form 
as
\beq
&&W_{\mu\nu}(p,q,P_h)=
\int{dz\over z^2}D(z)
\int {dx_1 \over x_1}\int{dx_2\over x_2}\nonumber\\
&&\qquad\qquad\qquad\times\left[
\left. 
{\partial S^{abc}_{\mu\nu;\rho\tau\lambda}(k_1,k_2,q,p_c)p^\lambda
\over \partial k_2^\sigma}\right|_{k_i=x_ip}\right]^{\rm pole}
\omega^\rho_{\ \alpha}\,\omega^\tau_{\ \beta}\,\omega^\sigma_{\ \gamma}\,
{\cal M}^{\alpha\beta\gamma}_{F,abc}(x_1,x_2), 
\label{wfinal}
\eeq
up to the twist-3 accuracy in the collinear factorization
with $k_1=x_1 p$ and $k_2 =x_2 p$,
where 
$\omega^\rho_{\ \,\alpha}=g^\rho_{\ \,\alpha}-p^\rho n_\alpha$ and  
${\cal M}^{\alpha\beta\gamma}_{F,abc}(x_1,x_2)$ denotes
the three-gluon lightcone correlation functions 
defined in terms of the gluon field-strength tensors, as
\beq
{\cal M}^{\alpha\beta\gamma}_{F,abc}(x_1,x_2)
&& =-gi^3\int{d\lambda\over 2\pi}\int{d\mu\over 2\pi}e^{i\lambda x_1}
e^{i\mu(x_2-x_1)}
\la pS|F_b^{\beta n}(0)F_c^{\gamma n}(\mu n)F_a^{\alpha n}(\lambda n)
|pS\ra
\nonumber\\
&&=\frac{3}{40}d^{abc}O^{\alpha\beta\gamma}(x_1,x_2)
-\frac{i}{24}f^{abc}N^{\alpha\beta\gamma}(x_1,x_2) ,
\label{MFabc}
\eeq
with $O^{\alpha\beta\gamma}(x_1,x_2)$ and 
$N^{\alpha\beta\gamma}(x_1,x_2)$ given in (\ref{3gluonO}) and (\ref{3gluonN}).
We introduced the notation $\left[ \cdots \right]^{\rm pole}$, where
$\left[ S^{abc}_{\mu\nu;\rho\tau\lambda}(k_1,k_2,q,p_c)\right]^{\rm pole}$ implies that 
the bared propagator arising in Fig.~2 (and its mirror diagrams) should be replaced by its pole part
$\propto \delta\left((p_c +k_1-k_2)^2 -m_c^2 \right)$
(or $\delta\left((p_c +k_2-k_1)^2 -m_c^2 \right)$), 
with 
$p_c$ 
the momentum of the $c$-quark fragmenting into the $D$-meson, $p_c^2=m_c^2$;
this fixes a momentum fraction at $x_1=x_2$ in the collinear limit $k_i=x_i p$ in (\ref{wfinal}).
$p_c$ 
is parameterized by the momentum fraction $z$ associated with the fragmentation function $D(z)$, as
\beq
p_c^\mu &&=\frac{1}{z}P_h^\mu +\frac{1}{2p\cdot P_h}\left(m_c^2z - \frac{m_h^2}{z} \right)p^\mu 
\nonumber\\
&&= {\zhat Q \over 2}\left( 1 + {q_T^2\over Q^2}+ {m_c^2\over \zhat^2Q^2},{2 q_T\over Q}\cos\chi,
{2 q_T\over Q}\sin\chi,
-1+{q_T^2\over Q^2}+{m_c^2\over \zhat^2Q^2}\right)  ,
\label{charmmom}
\eeq 
where $\zhat =z_f/z$, and the second line shows the explicit form in the hadron 
frame with (\ref{pmu}), (\ref{Dmomem}).  
The summation over the $c$ and $\bar{c}$ quark contributions, as well as the corresponding flavor index
on $D(z)$ and $S^{abc}_{\mu\nu;\rho\tau\lambda}(k_1,k_2,q,p_c)$, is implicit in (\ref{wfinal}).
Actually, many terms of twist-3, other than those in (\ref{wfinal}), are generated by the 
collinear expansion of the 
diagram of Fig.~1, and appear to be gauge-noninvariant.
We emphasize that all these gauge-noninvariant twist-3 terms 
either cancel or vanish
in the cross section, as demonstrated in \cite{BKTY10} by the use of Ward identities.

\begin{figure}[t!]
\begin{center}
\epsfig{figure=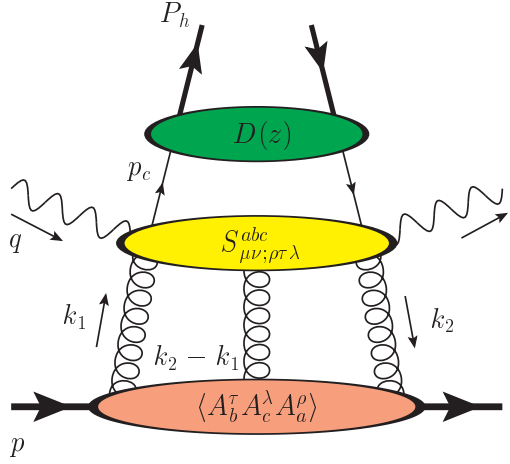,width=0.43\textwidth}
\end{center}
\caption{
Generic diagram giving rise to the twist-3 contribution to the 
hadronic tensor of $ep^\uparrow\to eDX$
induced by the gluonic effect in the nucleon.  It is
decomposed into the nucleon matrix element (lower blob), $D$-meson matrix element (upper blob),
and the partonic hard scattering part by the virtual photon (middle blob). 
\label{fig1}
}
\end{figure}

\begin{figure}[t!]
\begin{center}
\epsfig{figure=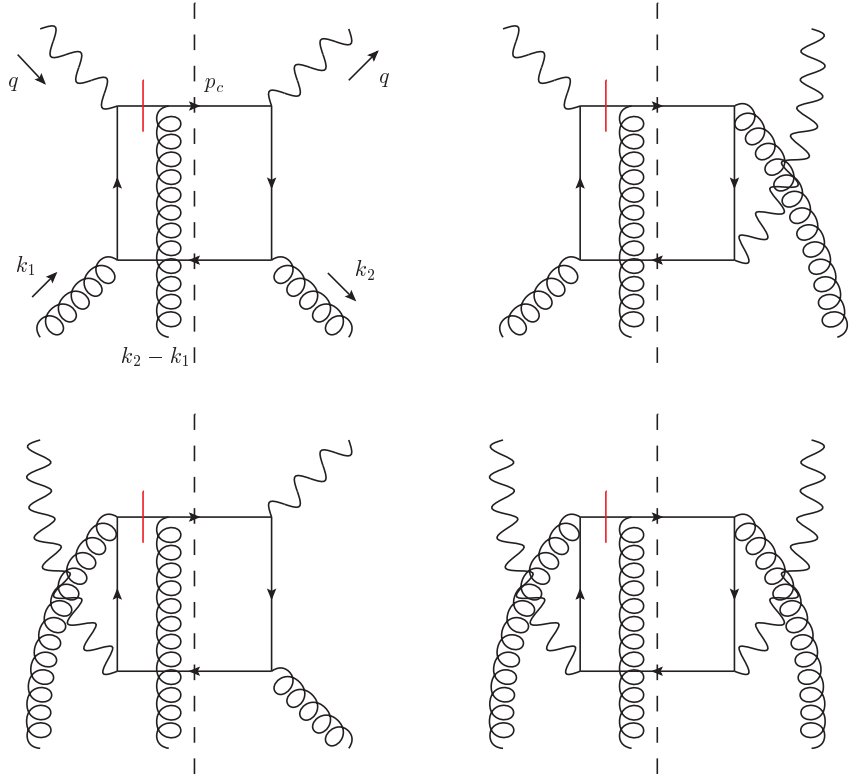,width=0.7\textwidth}
\end{center}
\caption{
Feynman diagrams for the partonic hard part in Fig.~1, representing the 
photon-gluon fusion subprocesses that give rise to the pole contribution for $ep^\uparrow\to eDX$
in the leading order with respect to the QCD coupling constant. 
The short bar on the internal $c$-quark line
indicates that the pole part is to be taken from that propagator.  
In the text, momenta are assigned as shown in the upper-left diagram, where
$p_c$ denotes the momentum of the $c$-quark fragmenting into the $D$-meson
in the final state.  The mirror diagrams also contribute.  
\label{fig2}
}
\end{figure}

\subsection{Calculation of $L_{\mu\nu}W^{\mu\nu}$}

To calculate the contraction $L^{\mu\nu}(\ell, \ell')W_{\mu\nu}(p,q,P_h)$ in (\ref{diffsigma}), 
we introduce the following four vectors which are orthogonal to
each other, similarly as in \,\cite{BKTY10}:
\beq
T^\mu &=&{1\over Q}\left(q^\mu + 2x_{bj}p^\mu\right),\nonumber\\
X^\mu &=&{1\over q_T}\left\{ {P_h^\mu \over z_f} -q^\mu -\left(
1+{q_T^2+m_h^2/z_f^2 \over Q^2}\right)x_{bj} p^\mu\right\},\nonumber\\
Y^\mu &=& \epsilon^{\mu\nu\rho\sigma}Z_\nu X_\rho T_\sigma ,\nonumber\\
Z^\mu &=& -{q^\mu \over Q}. 
\label{txyz}
\eeq
In the hadron frame
specified by (\ref{pmu}) and (\ref{Dmomem}), these vectors become
$T^\mu=(1,0,0,0)$, $X^\mu=(0,\cos\chi,\sin\chi,0)$, $Y^\mu=(0,-\sin\chi,\cos\chi,0)$, $Z^\mu=(0,0,0,1)$,
so that
\beq
(0,1,0,0)&&=\cos\chi X^\mu -\sin\chi Y^\mu ,
\nonumber\\
(0,0,1,0)&&=\sin\chi X^\mu + \cos\chi Y^\mu,
\label{basisbasis}
\eeq
In the present case, $W^{\mu\nu}$ can be expanded in terms of
the following six independent tensors\,\cite{BKTY10}: 
\beq
&&{\cal V}_1^{\mu\nu}=X^\mu X^\nu + Y^\mu Y^\nu,\qquad
{\cal V}_2^{\mu\nu}=g^{\mu\nu} + Z^\mu Z^\nu,\nonumber\\
&&{\cal V}_3^{\mu\nu}=T^\mu X^\nu + X^\mu T^\nu,\qquad
{\cal V}_4^{\mu\nu}=X^\mu X^\nu - Y^\mu Y^\nu,\nonumber\\
&&{\cal V}_8^{\mu\nu}=T^\mu Y^\nu + Y^\mu T^\nu,\qquad
{\cal V}_9^{\mu\nu}=X^\mu Y^\nu + Y^\mu X^\nu. 
\eeq
We also introduce
the inverse tensors $\widetilde{\cal V}_k^{\mu\nu}$ 
for the above ${\cal V}_k^{\mu\nu}$: 
\beq
&&\widetilde{{\cal V}}_1^{\mu\nu}={1\over 2}(2T^\mu T^\nu
+X^\mu X^\nu + Y^\mu Y^\nu),\qquad
\widetilde{{\cal V}}_2^{\mu\nu}=T^\mu T^\nu,\nonumber\\
&&\widetilde{{\cal V}}_3^{\mu\nu}=-{1\over 2}(T^\mu X^\nu + X^\mu T^\nu),\qquad
\widetilde{{\cal V}}_4^{\mu\nu}={1\over 2}(X^\mu X^\nu - Y^\mu Y^\nu),\nonumber\\
&&\widetilde{{\cal V}}_8^{\mu\nu}={-1\over 2}(T^\mu Y^\nu + Y^\mu T^\nu),\qquad
\widetilde{{\cal V}}_9^{\mu\nu}={1\over 2}(X^\mu Y^\nu + Y^\mu X^\nu).  
\eeq
With these definitions, one has
\beq
L_{\mu\nu}W^{\mu\nu}=  \sum_{k=1,\cdots,4,8,9} \left[ L_{\mu\nu}{\cal V}_k^{\mu\nu} \right]
\left[W_{\rho\sigma}\widetilde{\cal V}_k^{\rho\sigma}\right]
\equiv Q^2\sum_{k=1,\cdots,4,8,9} {\cal A}_k(\phi-\chi)
\left[W_{\rho\tau}\widetilde{\cal V}_k^{\rho\tau}\right], 
\label{LW}
\eeq
where ${\cal A}_k(\phi-\chi) \equiv L_{\mu\nu}{\cal V}_k^{\mu\nu}/Q^2$
parameterize the dependence of the cross section
on the azimuthal angle $\phi$ of the lepton plane 
relative to the hadron plane (see (\ref{eq2.lepton}), (\ref{Dmomem})),
and depend on $\phi$ and $\chi$ through $\phi-\chi$, with
\beq
{\cal A}_1(\phi)&=&1+\cosh^2\psi,\nonumber\\
{\cal A}_2(\phi)&=&-2,\nonumber\\
{\cal A}_3(\phi)&=&-\cos\phi\sinh 2\psi,\nonumber\\
{\cal A}_4(\phi)&=&\cos 2\phi\sinh^2\psi,\nonumber\\
{\cal A}_8(\phi)&=&-\sin\phi\sinh 2\psi,\nonumber\\
{\cal A}_9(\phi)&=&\sin 2\phi\sinh^2\psi. 
\label{Ak}
\eeq
By the expansion (\ref{LW}), 
the cross section for $ep^\uparrow\to eDX$
consists of the five structure functions associated with
${\cal A}_{1,2}$, ${\cal A}_3$, ${\cal A}_4$, ${\cal A}_8$
and ${\cal A}_9$, respectively, which have
different dependences
on the azimuthal angle $\phi$.

\section{Master formula for three-gluon contribution}

\subsection{Connection between the twist-3 hard part and the $\gamma^*g\to c\bar{c}$ scattering} 

To obtain the twist-3 cross section based on (\ref{wfinal}), 
one has to calculate the corresponding hard part as the derivative,
$\partial S^{abc}_{\mu\nu;\alpha\beta\lambda}(k_1,k_2,q,p_c)p^\lambda/\partial k_2^\gamma$, for
the contributions from the diagrams in Fig.~2,
followed by the collinear limit $k_i\to x_ip$.
We note that many building blocks in the diagrams in Fig.~2 depend on $k_2$
due to the participation of three external gluons with the momenta $k_1$, $k_2$ and $k_2-k_1$,
so that taking the derivative with respect to $k_2^\gamma$
produces many terms in the intermediate step and is quite complicated. 
As will be shown below, the above hard part for the twist-3 cross section is 
connected to the hard part with only the two external gluons,
representing the ``twist-2 level'' partonic scattering, $\gamma^*g\to c\bar{c}$.
This implies, in particular, some relevant contribution to the hard part for the twist-3 cross section 
is completely determined by the hard part for 
the twist-2 unpolarized process $ep \to eDX$, which is associated with the gluon density-distribution
in the nucleon,
and, similarly,
the spin-dependent contributions in the partonic scattering $\gamma^*g\to c\bar{c}$ completely determine 
the remaining hard part for the twist-3 cross section.

In order to prove this, 
we proceed similarly as in the proof in \cite{KT071,KT072}
for the master formula associated with the quark-gluon correlation,
but we discuss each step of our proof in detail
because it involves extensions 
for the case not only with the three-gluon correlation, but also with the nonzero 
quark-mass, compared to the
massless quark case treated in \cite{KT071,KT072}.
We first note that the hard part shown by the diagrams in Fig.~2 
has the structure obtained by attaching an extra gluon-line to the $c$-quark line 
in the diagrams in Fig.~3,
which show the leading-order contribution for the $\gamma^*g\to c\bar{c}$ scattering
with the $c$-quark 
fragmenting into the final $D$-meson.
We denote the sum of the contributions of the diagrams in Fig.~3 as
$S_{\mu\nu;\alpha\beta}^{(2)ab}(xp,q,p_c)$,
where the Lorentz indices $\alpha$ and $\beta$, as well as the color
indices $a$ and $b$, are associated with the external gluon lines that have the momentum $xp$.
$S_{\mu\nu;\alpha\beta}^{(2)aa}(xp,q,p_c)/8$ with the color indices averaged over
represents the hard part for the twist-2 cross sections.
In particular, the twist-2 unpolarized cross section
in (\ref{diffsigma}) is given by the contribution to the hadronic tensor,
\beq
W_{\mu\nu}^U(p,q,P_h)=
\int{dz\over z^2}D(z)
\int {dx \over x}
 S^{(2)ab}_{\mu\nu;\alpha\beta}(xp,q,p_c)
{\cal G}^{\alpha\beta}_{ab}(x),
\label{unpol}
\eeq
where ${\cal G}^{\alpha\beta}_{ab}(x)$ is the light-cone correlation function 
of the gluon's field strength tensors
in the nucleon, defined as 
\beq
{\cal G}^{\alpha\beta}_{ab}(x)={1\over x}\int{d\lambda\over 2\pi}e^{i\lambda x}
\la pS| F^{\beta n}_b(0)F^{\alpha n}_a(\lambda n)|pS\ra 
= -{1\over 2}g_\perp^{\alpha\beta}\times{1\over 8}\delta_{ab}G(x)+\cdots,
\label{gxgx}
\eeq
with the unpolarized gluon density $G(x)$, 
and $g_\perp^{\alpha\beta}=g^{\alpha\beta}-p^\alpha n^\beta -n^\alpha p^\beta$.
From Fig.~3, 
$S_{\mu\nu;\alpha\beta}^{(2)ab}(xp,q,p_c)$ 
is obtained as
\beq
S_{\mu\nu;\alpha\beta}^{(2)ab}(xp,q,p_c)
= {\rm Tr}\left[ \bar{\cal F}_\beta^b(xp,q,p_c)\left(\pslash_c+m_c\right)
{\cal F}_\alpha^a(xp,q,p_c) {\cal D}(xp+q-p_c) \right], 
\label{Stwist2}
\eeq
where ${\rm Tr}[\cdots]$ indicates the trace over 
both Dirac and color indices,
$\left(\pslash_c+m_c\right)$ with (\ref{charmmom}) is the projection matrix for the $D$-meson
fragmentation function, and the factor,
\begin{equation}
{\cal D}(k)\equiv 2\pi (\kslash -m_c)\delta(k^2-m_c^2),
\label{dk}
\end{equation}
is associated with the final-state cut for the unobserved $\bar{c}$-quark 
with the momentum $k$ flowing from the left to the right of the cut.  
The factor ${\cal F}_\alpha^a(xp,q,p_c)$ is the $\gamma g cc$-vertex function in the left of the cut,
containing 
the 
photon-quark and the gluon-quark vertices linked by the quark propagator; here,
the Lorentz and color indices, $\alpha$ and $a$, are associated with the external gluon, while
the Lorentz index $\mu$ for the virtual photon 
is suppressed for simplicity.  
The similar factor in the right of the cut can be written as
$\bar{\cal F}_\beta^b(xp,q,p_c) \equiv \gamma^0 [{{\cal F}_\beta^b(xp,q,p_c)}]^\dagger \gamma^0$.

\begin{figure}[t!]
\begin{center}
\epsfig{figure=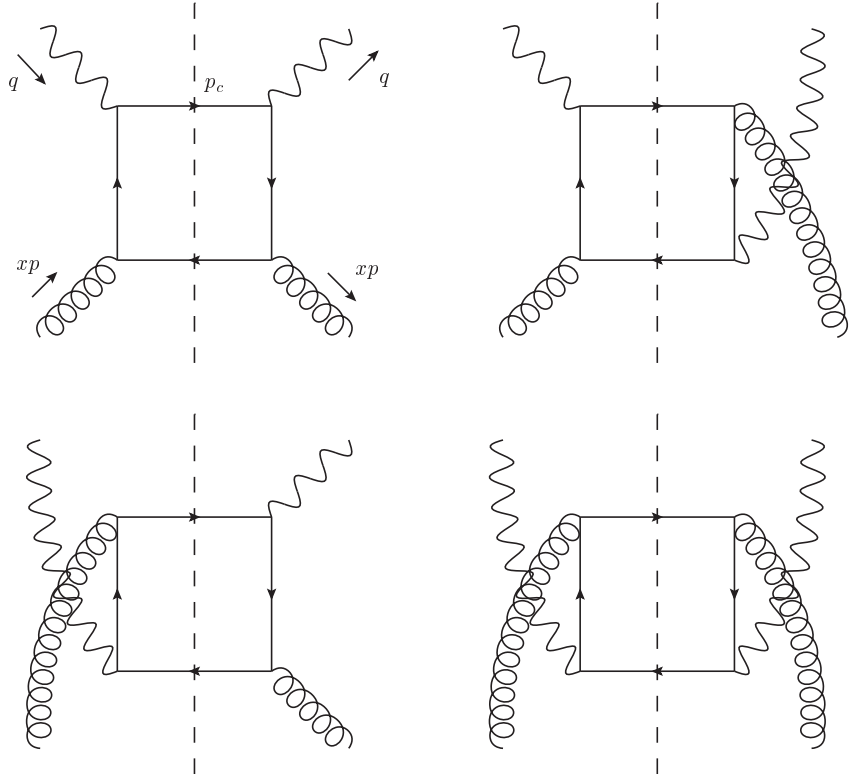,width=0.7\textwidth}
\end{center}
\caption{Leading-order diagrams for the $\gamma^*g\to c\bar{c}$ scattering cross section.
\label{fig3}
}
\end{figure}

With the replacement $xp \rightarrow k_1$ in (\ref{Stwist2}),
where $k_1$ has nonzero transverse components,
$S_{\mu\nu;\alpha\beta}^{(2)ab}(xp,q,p_c)$ may be 
extended to $S_{\mu\nu;\alpha\beta}^{(2)ab}(k_1,q,p_c)$.
$S^{abc}_{\mu\nu;\alpha\beta\lambda}(k_1,k_2,q,p_c)p^\lambda$ arising in (\ref{wfinal})
is obtained by attaching an additional gluon to $S_{\mu\nu;\alpha\beta}^{(2)ab}(k_1,q,p_c)$,
in particular, to the $c$-quark line fragmenting into the $D$ meson,
where the gluon carries the momentum $k_2 -k_1$, the 
polarization associated with $p^\lambda$, and the color index $c$.
When the extra gluon is attached in the left of the cut as in Fig.~2,
we obtain
\begin{equation}
S_{L,\alpha\beta}^{abc}(k_1,k_2,q,p_c)
= {\rm Tr}\left[ \bar{\cal F}_\beta^b(k_2,q,p_c)t^cL(k_1-k_2+p_c)
{\cal F}_\alpha^a(k_1,q,k_1-k_2+p_c){\cal D}(k_2+q-p_c)\right], 
\label{Sleft}
\end{equation}
where 
\beq
L(k_1-k_2+p_c)=\left(\pslash_c+m_c\right)\pslash {-1\over \kslash_1-\kslash_2+\pslash_c -m_c +i\epsilon},
\label{LL}
\eeq
and $t^c$ is the color matrix for the quark-gluon vertex
(the coupling constant $g$ for the quark-gluon vertex is absorbed in  
${\cal M}^{\alpha\beta\gamma}_{F,abc}(x_1,x_2)$ of (\ref{MFabc})).
Likewise, 
when the extra gluon is attached in the right of the cut as in the mirror diagrams of Fig.~2,
we obtain
\begin{equation}
S_{R,\alpha\beta}^{abc}(k_1,k_2,q,p_c)
={\rm Tr}\left[ \bar{\cal F}_\beta^b(k_2,q,k_2-k_1+p_c) t^c R(k_2-k_1+p_c)
{\cal F}_\alpha^a(k_1,q,p_c){\cal D}(k_1+q-p_c)\right], 
\label{Sright}
\end{equation}
with
\beq
R(k_2-k_1+p_c)= 
{-1\over \kslash_2-\kslash_1 + \pslash_c-m_c -i\epsilon} \pslash \left(\pslash_c+m_c\right)
= \gamma^0 L^\dagger (k_2-k_1+p_c) \gamma^0, 
\label{RR}
\eeq
so that 
\begin{equation}
S_{R,\alpha\beta}^{abc}(k_1,k_2,q,p_c)={S_{L, \beta\alpha}^{bac}}^*(k_2,k_1,q,p_c) .
\label{SRL}
\end{equation}
Thus, $S_{\mu\nu;\alpha\beta\lambda}^{abc}(k_1,k_2,q,p_c)p^\lambda$
can be obtained as the sum of
(\ref{Sleft}) and (\ref{Sright}): 
\beq
S_{\mu\nu;\alpha\beta\lambda}^{abc}(k_1,k_2,q,p_c)p^\lambda=
S_{L,\alpha\beta}^{abc}(k_1,k_2,q,p_c)+S_{R,\alpha\beta}^{abc}(k_1,k_2,q,p_c).  
\label{Stwist3}
\eeq
Comparing (\ref{Stwist2}) with (\ref{Stwist3}), 
we see that $S_{\mu\nu;\alpha\beta\lambda}^{abc}(k_1,k_2,q,p_c)p^\lambda$ is obtained from 
$S_{\mu\nu;\alpha\beta}^{(2)ab}(k_2,q,p_c)$ by
the formal replacements, $\left(\pslash_c+m_c\right)\rightarrow t^cL(k_1-k_2+p_c)$ 
and $\left(\pslash_c+m_c\right)\rightarrow t^cR(k_2-k_1+p_c)$,
together with the appropriate momentum-shifts in the remaining factors. 

We calculate 
$\partial S^{abc}_{\mu\nu;\alpha\beta\lambda} (k_1,k_2,q,p_c)
p^\lambda\left. /\partial k_2^\gamma
\right|_{k_i=x_ip}$ in (\ref{wfinal}) based on (\ref{Sleft})-(\ref{Stwist3}),
keeping their relevant structure that manifests 
the above-mentioned correspondence with $S_{\mu\nu;\alpha\beta}^{(2)ab}(k_2,q,p_c)$.
For this purpose, we need the collinear limit 
of (\ref{LL}) and (\ref{RR}),
\beq
L\left((x_1-x_2)p+p_c\right)= -R\left((x_2-x_1)p+p_c\right)= 
\left(\pslash_c+m_c\right){-1\over x_1-x_2+i\epsilon}, 
\label{LR}
\eeq
and of
their derivatives,
\beq
&&
\!\!\!\!\!\!\!\!
\left.{\partial L(k_1-k_2+p_c) \over \partial k_2^\alpha } \right|_{k_i=x_ip}
={\left( \pslash_c+m_c\right)\pslash\gamma_\alpha \over 2p\cdot p_c}{1\over x_1-x_2+i\epsilon}
-{p_{c\alpha}\left( \pslash_c+m_c\right)\over p\cdot p_c}{1\over (x_1-x_2+i\epsilon)^2},\label{LRderiv0}\\
&&
\!\!\!\!\!\!\!\!
\left.{\partial R(k_2-k_1+p_c) \over \partial k_2^\alpha } \right|_{k_i=x_ip}
={\gamma_\alpha \pslash \left( \pslash_c+m_c\right)\over 2p\cdot p_c}{1\over x_1-x_2+i\epsilon}
+{p_{c\alpha}\left( \pslash_c+m_c\right)\over p\cdot p_c}{1\over (x_1-x_2+i\epsilon)^2}. 
\label{LRderiv}
\eeq
In these relations~(\ref{LR})-(\ref{LRderiv}), the poles relevant to $\left[ \cdots \right]^{\rm pole}$
in (\ref{wfinal}) are unveiled; it is straightforward to see
that the $\gamma g cc$-vertex functions and their derivatives do not produce
the pole contributions in the $k_i\to x_ip$ limit with $x_i >0$.
Based on the pole structures in (\ref{LR})-(\ref{LRderiv}), 
we evaluate the pole contribution in (\ref{wfinal}) as
\begin{equation}
\left[
\left.{\partial S^{abc}_{\mu\nu;\alpha\beta\lambda}(k_1,k_2,q,p_c)p^\lambda \over \partial k_2^\gamma}\right|_{k_i=x_ip}
\right]^{\rm pole}=\left[
\left.{\partial S^{abc}_{\mu\nu;\alpha\beta\lambda}(k_1,k_2,q,p_c)p^\lambda \over \partial k_2^\gamma}\right|_{k_i=x_ip}
\right]^{\rm (i)+(ii)+(iii)},
\label{sumsum}
\end{equation}
decomposing it into three parts (i)-(iii), where
(i) denotes the pole contributions from the second term in (\ref{LRderiv0}) and (\ref{LRderiv}), 
(ii) denotes those 
from the first term in (\ref{LRderiv0}) and (\ref{LRderiv}), and (iii) denotes
the remaining pole contributions due to (\ref{LR}). 
We obtain
\beq
&&\left[
\left.{\partial S^{abc}_{\mu\nu;\alpha\beta\lambda}(k_1,k_2,q,p_c)p^\lambda \over \partial k_2^\gamma}\right|_{k_i=x_ip}
\right]^{\rm (i)}={p_{c\gamma}\over p\cdot p_c}
\left[{1\over (x_1-x_2+i\epsilon)^2}\right]^{\rm pole}
\nonumber\\
&&\qquad\qquad\times{\rm Tr}\left[-\bar{\cal F}_\beta^b(x_2p,q,p_c)t^c\left(\pslash_c+m_c\right)
{\cal F}_\alpha^a(x_1p,q,(x_1-x_2)p+p_c){\cal D}(x_2p+q-p_c)\right.\nonumber\\
&&\left.\qquad\qquad
+\bar{\cal F}_\beta^b(x_2p,q,(x_2-x_1)p+p_c)t^c\left(\pslash_c+m_c\right)
{\cal F}_\alpha^a(x_1p,q,p_c){\cal D}(x_1p+q-p_c)\right]
\nonumber\\
&&\qquad\quad={p_{c\gamma}\over p\cdot p_c}
\left[{-1\over x_1-x_2+i\epsilon}\right]^{\rm pole}
\nonumber\\
&&\qquad\qquad\times {\rm Tr}\left[\bar{\cal F}_\beta^b(x_1p,q,p_c)t^c\left(\pslash_c+m_c\right)
\left(
p^\mu{\partial \over \partial p_c^\mu}{\cal F}_\alpha^a(x_1p,q,p_c)\right) {\cal D}(x_1p+q-p_c)
\right.\nonumber\\
&&\left.\qquad\qquad
+\left(
p^\mu {\partial \over \partial p_c^\mu}\bar{\cal F}_\beta^b(x_1p,q,p_c)\right)
t^c\left(\pslash_c+m_c\right)
{\cal F}_\alpha^a(x_1p,q,p_c){\cal D}(x_1p+q-p_c)
\right.\nonumber\\
&&\left.\qquad\qquad
+\bar{\cal F}_\beta^b(x_1p,q,p_c)t^c\left(\pslash_c+m_c\right)
{\cal F}_\alpha^a(x_1p,q,p_c)\left(
p^\mu{\partial \over \partial p_c^\mu}{\cal D}(x_1p+q-p_c)\right) 
\right]\nonumber\\
&&\qquad\quad={p_{c\gamma}\over p\cdot p_c}\left[{-1\over x_1-x_2+i\epsilon}\right]^{\rm pole}\nonumber\\
&&\qquad\qquad\times
{\rm Tr}\left[
p^\mu {\partial \over \partial p_c^\mu}
\bar{\cal F}_\beta^b(x_1p,q,p_c)
t^c\left(\pslash_c+m_c\right)
{\cal F}_\alpha^a(x_1p,q,p_c){\cal D}(x_1p+q-p_c)
\right.\nonumber\\
&&\qquad\qquad\qquad
-\bar{\cal F}_\beta^b(x_1p,q,p_c)
t^c\ \! \pslash\ \!
{\cal F}_\alpha^a(x_1p,q,p_c){\cal D}(x_1p+q-p_c)\biggr].
\label{doublepole}
\eeq
In the second equality in (\ref{doublepole}), we have performed the Taylor expansion
with respect to $x_2$ around $x_1$
for the contributions inside the trace ${\rm Tr}$, 
and the partial derivative with respect to $p_c$ implies the shorthand notation of
\begin{equation}
\frac{\partial}{\partial p_c^\mu}f(p_c) \equiv \left. \frac{\partial}{\partial r^\mu}f(r) \right|_{r\rightarrow p_c},
\label{shorthand}
\end{equation}
for a function $f(r)$ of a four-vector $r_\mu$.
The leading-order (zeroth-order) term in the Taylor expansion would give rise 
to the double pole contribution to (\ref{doublepole}),
but the corresponding contributions generated from the two terms in the RHS of (\ref{Stwist3})
cancel; this type of cancellation may be considered as 
a result of gauge invariance~\cite{KT071}. On the other hand, 
the second and higher-order terms in the expansion
give the vanishing contribution to (\ref{doublepole}) when combined with 
the factor $\left[1/ (x_1-x_2+i\epsilon)^2\right]^{\rm pole}$.
Similarly, for the contribution labeled by (ii) above, we easily obtain
\beq
&&\!\!\!\!\!\!\!\!\!\!\!\!\!\!
\left[
\left.{\partial S^{abc}_{\mu\nu;\alpha\beta\lambda}(k_1,k_2,q,p_c)p^\lambda 
\over \partial k_2^\gamma}\right|_{k_i=x_ip}
\right]^{\rm (ii)}\nonumber\\
&&\!\!\!\!\!\!\!\!\!\!\!\!\!\!=
\left[{1\over x_1-x_2+i\epsilon}\right]^{\rm pole}
{\rm Tr} \left[\bar{\cal F}_\beta^b(x_1p,q,p_c)
t^c\left(\gamma_\gamma-{p_{c\gamma}\over p\cdot p_c}\pslash\right)
{\cal F}_\alpha^a(x_1p,q,p_c){\cal D}(x_1p+q-p_c) \right]. 
\label{singlepole}
\eeq
The contribution labeled by (iii) 
can be expressed as
\beq
&&\left[\left.{\partial S^{abc}_{\mu\nu;\alpha\beta\lambda}
(k_1,k_2,q,p_c)p^\lambda \over \partial k_2^\gamma}\right|_{k_i=x_ip}
\right]^{\rm (iii)}\nonumber\\
&&\quad= \left[{1\over x_1-x_2+i\epsilon}\right]^{\rm pole}
{\rm Tr}\left[
{\partial \over \partial p_c^\gamma}
\bar{\cal F}_\beta^b(x_1p,q,p_c)
t^c\left(\pslash_c+m_c\right)
{\cal F}_\alpha^a(x_1p,q,p_c){\cal D}(x_1p+q-p_c)
\right.\nonumber\\
&&\qquad\qquad\qquad\qquad
-\bar{\cal F}_\beta^b(x_1p,q,p_c)
t^c\ \! \gamma_\gamma\ \!
{\cal F}_\alpha^a(x_1p,q,p_c){\cal D}(x_1p+q-p_c)\biggr],
\label{rest}
\eeq
where
we have replaced the derivative ${\partial/\partial k_2^\gamma}$ by the relevant
${\partial / \partial p_c^\gamma}$, and have set $x_2=x_1$ inside the  ${\rm Tr}$
due to the presence of the factor $\left[1/ (x_1-x_2+i\epsilon)\right]^{\rm pole}$.
The sum of the results (\ref{doublepole})-(\ref{rest})
yields the compact form for (\ref{sumsum}), 
\begin{equation}
\left[\left.{\partial S^{abc}_{\mu\nu;\alpha\beta\lambda}
(k_1,k_2,q,p_c)p^\lambda \over \partial k_2^\gamma}\right|_{k_i=x_ip}
\right]^{\rm pole}=
-i\pi\delta(x_1-x_2)
\left(
{\partial \over \partial p_c^\gamma} - {p_{c\gamma}p^\mu \over p\cdot p_c}
{\partial\over \partial p_c^\mu}
\right)
\widetilde{S}_{\mu\nu;\alpha\beta}^{abc}(x_1p,q,p_c), 
\label{master1}
\end{equation}
where
\beq
\widetilde{S}_{\mu\nu;\alpha\beta}^{abc}(x_1p,q,p_c)=
{\rm Tr}\left[\bar{\cal F}_\beta^b(x_1p,q,p_c)
t^c\left(\pslash_c+m_c\right)
{\cal F}_\alpha^a(x_1p,q,p_c){\cal D}(x_1p+q-p_c)\right].
\label{Smaster}
\eeq
Noting that the RHS of (\ref{master1}) vanishes when contracted by $p^\gamma$,
(\ref{master1}) may be calculated as
\beq
\left[\left.{\partial S^{abc}_{\mu\nu;\alpha\beta\lambda}(k_1,k_2,q,p_c)p^\lambda 
\over \partial k_2^\gamma}\right|_{k_i=x_ip}
\right]^{\rm pole}= 
-i\pi\delta(x_1-x_2)
{d \over d p_c^\gamma} \widetilde{S}_{\mu\nu;\alpha\beta}^{abc}(x_1p,q,p_c), 
\label{master2}
\eeq
where the derivative is to be taken under
the on-shell condition $p_c^2=m_c^2$,
regarding 
$p_c^+$ as a variable dependent on $p_c^-$ and $p_c^{1,2}$ 
as in (\ref{charmmom}).
Comparing (\ref{Smaster}) and (\ref{Stwist2}), 
$\widetilde{S}_{\mu\nu;\alpha\beta}^{abc}(x_1p,q,p_c)$ 
is the same as $S_{\mu\nu;\alpha\beta}^{(2)ab}(x_1p,q,p_c)$, except that
the former has an extra insertion of the color matrix $t^c$.  
The relations (\ref{master1}) and (\ref{master2}) 
clearly indicate that
the three-gluon contribution to the twist-3 cross section for $ep^\uparrow \to eDX$ (see (\ref{wfinal}))
can be derived from the cross sections
for the $\gamma^*g\to c\bar{c}$ scattering, and 
we find that those relations, obtained for the three-gluon contribution and with a massive quark,
have the structure formally similar to the corresponding relations
derived in \cite{KT071,KT072}
for the quark-gluon contribution and with the massless quarks.
Substituting (\ref{wfinal}), (\ref{master2}) into (\ref{diffsigma}),
we obtain the master formula, the main result of this paper,
which allows us to derive
the explicit form of the whole contribution to the twist-3 SSA in $ep^\uparrow \to eDX$,
as we demonstrate in the next section.

\subsection{Calculation of the twist-3 cross section based on master formula}

We are now in a position to carry out the 
derivative arising in (\ref{master2}).
In the formula (\ref{wfinal}) with (\ref{MFabc}) and (\ref{master2}),
it is sufficient to consider the corresponding derivative 
for the color-averaged components of
$\widetilde{S}^{abc}_{\mu\nu;\alpha\beta}(xp,q,p_c)$,
which coincide with 
the color-averaged component
of $S^{(2)ab}_{\mu\nu;\alpha\beta}(xp,q,p_c)$, as apparent
from the color structure in (\ref{Stwist2}), (\ref{Smaster}). 
We thus define
the color-averaged hard part $S^{(2)}_{\mu\nu;\alpha\beta}(xp,q,p_c)$ as the appropriate color contraction:
\begin{equation}
\frac{1}{8} \delta^{ab}S^{(2)ab}_{\mu\nu;\alpha\beta}(xp,q,p_c)\equiv 
{1\over 2}S^{(2)}_{\mu\nu;\alpha\beta}(xp,q,p_c).
\label{cav}
\end{equation}
Then, we have
\begin{equation}
{3\over 40}d^{abc}\widetilde{S}^{abc}_{\mu\nu;\alpha\beta}(xp,q,p_c)
={-i\over 24}f^{abc}\widetilde{S}^{abc}_{\mu\nu;\alpha\beta}(xp,q,p_c) = {1\over 4}S^{(2)}_{\mu\nu;\alpha\beta}(xp,q,p_c).
\label{cav2}
\end{equation}
Furthermore, the above derivative 
can be calculated most conveniently after taking the contraction of
$S^{(2)}_{\mu\nu;\alpha\beta}(xp,q,p_c)$ with the leptonic tensor $L^{\mu\nu}$,
as implied in (\ref{diffsigma}). 
According to (\ref{LW}), 
we introduce 
\beq
\widetilde{\cal V}_k^{\mu\nu}S^{(2)}_{\mu\nu;\alpha\beta}(xp,q,p_c)\equiv
2\pi{g^2\over \zhat Q^2}H^k_{\alpha\beta}(xp,q,p_c). 
\label{Hdef}
\eeq
Here, in the RHS, 
we separated the coupling constant, contained in the $\gamma g cc$-vertex function 
${\cal F}_\alpha^a(xp,q,p_c)$ in (\ref{Stwist2}),
as well as the factor $2\pi/(\zhat Q^2)$ contained in ${\cal D}(xp+q-p_c)$
for the unobserved final-state, such that (see (\ref{dk}))
\beq
2\pi\delta\left( (xp+q-p_c)^2)-m_c^2 \right)= 2\pi{1\over \zhat Q^2} 
\delta\left(
\frac{q_T^2}{Q^2}-\left(1-\frac{1}{\hat{x}}\right)\left(1-\frac{1}{\hat{z}}\right)
+\frac{m_c^2}{\hat{z}^2Q^2}\right),
\label{onshell}
\eeq
with $\xhat=x_{bj}/x$ and $\zhat=z_f/z$. 
Thus,
one obtains 
\beq
&&L^{\mu\nu}W_{\mu\nu}= {2\pi g^2\over z_f}{1\over 4}
\sum_{k=1,\cdots,4,8,9}\int{dz\over z}D(z)\int{dx\over x^2}(-i\pi)
{d\over dp_c^\gamma}\left[
{\cal A}_k(\phi-\chi)
H^k_{\alpha\beta}(xp,q,p_c)\right] 
\nonumber\\
&&\qquad\qquad\qquad\qquad\qquad\times\left(O^{\alpha\beta\gamma}(x,x)
+N^{\alpha\beta\gamma}(x,x)
\right),
\label{LWS}
\eeq
for the single-spin-dependent contribution,
where (see (\ref{3gluonO}) and (\ref{3gluonN}))
\beq
&&O^{\alpha\beta\gamma}(x,x)=2iM_N\left[
O(x,x)g_\perp^{\alpha\beta}\epsilon^{\gamma pnS}
+O(x,0)(g_\perp^{\beta\gamma}\epsilon^{\alpha pnS}
+g_\perp^{\gamma\alpha}\epsilon^{\beta pnS})\right],\nonumber\\
&&N^{\alpha\beta\gamma}(x,x)=2iM_N\left[
N(x,x)g_\perp^{\alpha\beta}\epsilon^{\gamma pnS}
-N(x,0)(g_\perp^{\beta\gamma}\epsilon^{\alpha pnS}
+g_\perp^{\gamma\alpha}\epsilon^{\beta pnS})\right],  
\label{3gluonxx}
\eeq
up to the irrelevant terms corresponding to the twist higher than three.  
We note that (\ref{LWS}) with (\ref{3gluonxx}) is described by the two kinds of partonic hard parts:
the hard part associated with $O(x,x)$ is same as that for $N(x,x)$, but
$O(x,0)$ as well as $N(x,0)$ accompanies the hard part of another kind.

The result~(\ref{LWS}) shows that only the derivative with respect to the transverse components, 
$(p_{c}^1,p_{c}^2)= {\mathbf{p}}_{c\perp}$, contributes
to the twist-3 cross section.
As noted below (\ref{master2}),
those components can be varied as independent variables
when performing the derivative,
and, based on the representation in (\ref{charmmom}),
the corresponding derivative may be performed 
through the magnitude $p_{c\perp}\equiv |{\mathbf{p}}_{c\perp}| =\zhat q_T$ and the azimuthal angle $\chi$ 
of the transverse components,
as
\beq
{\partial \over \partial p_c^1}=\cos\chi {\partial \over \partial p_{c\perp}}
-{\sin\chi\over p_{c\perp}}{\partial \over \partial \chi}\,,
\qquad\qquad 
{\partial \over \partial p_c^2}=\sin\chi {\partial \over \partial p_{c\perp}}
+{\cos\chi\over p_{c\perp}}{\partial \over \partial \chi}. 
\label{derivative}
\eeq
The formulae (\ref{Ak}) indicate that ${\partial / \partial p_{c\perp}}$ 
hits only
$H^k_{\alpha\beta}(xp,q,p_c)$ in (\ref{LWS}). 
Substituting (\ref{3gluonxx}) into (\ref{LWS}) and using 
$\epsilon^{1pnS}=\sin\Phi_S$ and $\epsilon^{2pnS}=-\cos\Phi_S$, the hard cross section for
$\{O(x,x),N(x,x)\}$ can be obtained as 
\beq
&&
{d\over dp_c^\gamma}\left\{
{\cal A}_k(\phi-\chi)H^k_{\alpha\beta}\right\}
g_\perp^{\alpha\beta}\epsilon^{\gamma pnS} 
\nonumber\\
&&\qquad ={\sin(\Phi_S-\chi) \over \zhat}{\cal A}_k(\phi-\chi){\partial H^k_{\alpha\beta}
g_\perp^{\alpha\beta} \over \partial q_T}
+{\cos(\Phi_S-\chi) \over\zhat q_T}{\partial {\cal A}_k(\phi-\chi) \over \partial \phi}
H^k_{\alpha\beta}
g_\perp^{\alpha\beta},
\label{XsecXX}
\eeq
where we have used the fact that the scalar 
function $H^k_{\alpha\beta}g_\perp^{\alpha\beta}$ does not depend on the angular variable $\chi$.  
Indeed, comparing with (\ref{unpol}), (\ref{gxgx}),
we remark that 
$-H^k_{\alpha\beta}g_\perp^{\alpha\beta}$ is
nothing but the twist-2 partonic part
for the unpolarized gluon density $G(x)$, 
and its explicit form is calculated in \cite{BKTY10}.
Accordingly, the partonic hard part for
$\{O(x,x),N(x,x)\}$ is completely determined from the knowledge on the twist-2 unpolarized
cross section.  
Using the relations
${\partial {\cal A}_3 \over \partial \phi}=-{\cal A}_8$
and ${\partial {\cal A}_4 \over \partial \phi}=-2{\cal A}_9$,
we find that
the terms associated with the azimuthal structures ${\cal A}_{8,9}$ arise from the second term of (\ref{XsecXX}),
accompanying $\cos(\Phi_S-\chi)$,
although such azimuthal structures were absent from the twist-2 unpolarized
cross section (\ref{unpol}), i.e., $H^k_{\alpha\beta}g_\perp^{\alpha\beta}= 0$ for
$k=8, 9$ (see (\ref{unpolresult}) below).   
These remarkable features revealed in (\ref{XsecXX})
have been observed similarly in the master formula for the
SGP contribution to the SSA induced by the quark-gluon correlation~\cite{KT071}.
We also remind that the result (\ref{XsecXX}) depends on the azimuthal angles through $\phi-\chi$ and 
$\Phi_S-\chi$, as noted in section 3.1. 
\footnote{
Since this dependence is obvious from the beginning,
one may alternatively consider the derivative in (\ref{LWS}) with (\ref{derivative}) 
at $\chi=0$ to get $\Phi_S$-dependence
and restore the $\chi$-dependence by the shift $\Phi_S\to \Phi_S-\chi$ to reach (\ref{XsecXX}),
which is much simpler.  In other words, one can calculate the cross section for $\chi\rightarrow 0$
by regarding $\chi$ as the small parameter
that allows us to take the derivative $d/dp_c^\gamma$ in 
(\ref{LWS}).  
This also applies to (\ref{XsecXO}) below. }

The hard part for $\{O(x,0),N(x,0)\}$ can be calculated similarly.
In this case, however, we encounter the non-scalar components of $H^k_{\alpha\beta}$,
which have the dependence on the angular variable $\chi$.
The $\chi$ dependence of this type was irrelevant for the cases discussed in \cite{KT071,KT072},
but gives the novel contribution in the master-formula approach for the present case.
In order to deal with those non-scalar components,
we write down the corresponding hard part explicitly as
\beq
&&\hspace{-0.5cm}
{d\over dp_c^\gamma}\left\{
{\cal A}_k(\phi-\chi)H^k_{\alpha\beta}\right\}
\left(g_\perp^{\beta\gamma}\epsilon^{\alpha pnS} + g_\perp^{\alpha\gamma}\epsilon^{\beta pnS}\right)
\nonumber\\
&&
=-2{\partial \over \partial p_c^1}{\cal A}_k(\phi-\chi)H^k_{11}\epsilon^{1pnS}
-{\partial\over \partial p_c^1}{\cal A}_k(\phi-\chi)(H^k_{12}+H^k_{21})
\epsilon^{2pnS}
\nonumber\\
&&-
{\partial\over \partial p_c^2}
{\cal A}_k(\phi-\chi)(H^k_{12}+H^k_{21})\epsilon^{1pnS}
-2 {\partial\over \partial p_c^2}
{\cal A}_k(\phi-\chi)H^k_{22}\epsilon^{2pnS}
\nonumber\\
&&=-2\sin\Phi_S
\left( \cos\chi{\partial\over \partial p_{c\perp}} -{\sin\chi\over p_{c\perp}}
{\partial \over \partial \chi}\right)
{\cal A}_k(\phi-\chi)H^k_{11}\nonumber\\
&&\quad +\cos\Phi_S \left( \cos\chi{\partial\over \partial p_{c\perp}} -{\sin\chi\over p_{c\perp}}
{\partial \over \partial \chi}\right) {\cal A}_k(\phi-\chi)
(H^k_{12}+H^k_{21})
\nonumber\\
&&\quad -\sin\Phi_S\left( \sin\chi {\partial \over \partial p_{c\perp}}+
{\cos\chi \over p_{c\perp}}{\partial\over \partial \chi}\right)
{\cal A}_k(\phi-\chi)(H^k_{12}+H^k_{21})\nonumber\\
&&\quad +2\cos\Phi_S
\left( \sin\chi {\partial \over \partial p_{c\perp}}+
{\cos\chi \over p_{c\perp}}{\partial\over \partial \chi}\right)
{\cal A}_k(\phi-\chi)H^k_{22},
\eeq
and thus we need the $\chi$ dependence of $H^k_{11}$, $H^k_{12}+H^k_{21}$ and $H^k_{22}$.
This dependence can be separated using (\ref{basisbasis}) like
$H^k_{12}=H^k_{\alpha\beta}\left(\cos\chi X^\alpha -\sin\chi Y^\alpha\right)
\left(\sin\chi X^\beta + \cos\chi Y^\beta\right)$; 
as apparent from the definition~(\ref{txyz}) for the basis vectors
constructed with the momenta that are associated with the hadron plane,
$H^k_{XX}\equiv H^k_{\mu\nu}X^\mu X^\nu$, $H^k_{XY}\equiv H^k_{\mu\nu}X^\mu Y^\nu$, {\it etc}, 
are unchanged by the rotation of the hadron plane around the $z$-axis, i.e., are independent of $\chi$.
After some algebra, one obtains, in the obvious notation,
\beq
&&\hspace{-0.5cm}
{d\over dp_c^\gamma}\left\{
{\cal A}_k(\phi-\chi)H^k_{\alpha\beta}\right\}
\left(g_\perp^{\beta\gamma}\epsilon^{\alpha pnS} + g_\perp^{\alpha\gamma}\epsilon^{\beta pnS}\right)
\nonumber\\
&&=
\sin(\Phi_S-\chi) \left[
{-2 {\cal A}_k(\phi-\chi) \over\zhat}{\partial H^k_{XX} \over \partial q_T} \right.\nonumber\\
&&\left.\qquad\qquad\qquad +{1\over \zhat q_T}\left( {\partial {\cal A}_k(\phi-\chi) \over \partial \phi}
H^k_{(XY+YX)}
+2{\cal A}_k(\phi-\chi)H^k_{(-XX+YY)}
\right)\right]  \nonumber\\
&&+
\cos(\Phi_S-\chi) \left[
{{\cal A}_k(\phi-\chi) \over \zhat} {\partial H^k_{(XY+YX)} \over \partial q_T}\right.\nonumber\\
&&\left. \qquad\qquad\qquad 
+{1\over \zhat q_T}\left(
-2{\partial {\cal A}_k(\phi-\chi) \over \partial \phi}H^k_{YY} 
+ 2{\cal A}_k(\phi-\chi) H^k_{(XY+YX)}\right)\right].
\label{XsecXO}
\eeq
To proceed further,
we define the partonic hard cross sections for the ``$\gamma^* g\to c \bar{c}$" scattering, 
separating the delta function of (\ref{onshell}) and
the electric charge $e_c=2/3$ of the $c$-quark 
from the hard parts arising in (\ref{XsecXX}), (\ref{XsecXO}):
\beq
&&-H^k_{\alpha\beta}g_\perp^{\alpha\beta}=e_c^2 \hat{\sigma}_k^U(Q,q_T,\xhat,\zhat) \delta\left(
\frac{q_T^2}{Q^2}-\left(1-\frac{1}{\hat{x}}\right)\left(1-\frac{1}{\hat{z}}\right)
+\frac{m_c^2}{\hat{z}^2Q^2}\right),
\nonumber\\
&&-H^k_{XX}=e_c^2 \hat{\sigma}_k^{XX}(Q,q_T,\xhat,\zhat)\delta\left(
\frac{q_T^2}{Q^2}-\left(1-\frac{1}{\hat{x}}\right)\left(1-\frac{1}{\hat{z}}\right)
+\frac{m_c^2}{\hat{z}^2Q^2}\right),
\nonumber\\
&&-H^k_{YY}=e_c^2 \hat{\sigma}_k^{YY}(Q,q_T,\xhat,\zhat)\delta\left(
\frac{q_T^2}{Q^2}-\left(1-\frac{1}{\hat{x}}\right)\left(1-\frac{1}{\hat{z}}\right)
+\frac{m_c^2}{\hat{z}^2Q^2}\right),
\nonumber\\
&&-H^k_{(XY+YX)}=e_c^2 \hat{\sigma}_k^{\{XY\}}(Q,q_T,\xhat,\zhat)
\delta\left(
\frac{q_T^2}{Q^2}-\left(1-\frac{1}{\hat{x}}\right)\left(1-\frac{1}{\hat{z}}\right)
+\frac{m_c^2}{\hat{z}^2Q^2}\right),
\label{Hsigma}
\eeq
where 
the partonic cross sections $\hat{\sigma}^j_k$ ($j=U, XX, YY, \{XY\}$)
are the functions of $Q$, $q_T$, $\xhat$ and $\zhat$.
When (\ref{XsecXX}) and (\ref{XsecXO}) using these forms are
inserted into (\ref{LWS}),
the derivative $\partial / \partial q_T$ hitting the delta function in (\ref{Hsigma}) 
can be treated by integration by parts with respect to $x$;
for example, the contribution from the first term in the RHS of (\ref{XsecXX}), convoluted with the
correlation function 
$O(x,x)$,
can be calculated as 
\beq
&&{1\over \zhat}{\partial \over \partial q_T}
\int{dx\over x^2} \hat{\sigma}_k^{U}(Q,q_T,\xhat,\zhat)O(x,x)
\delta\left(
\frac{q_T^2}{Q^2}-\left(1-\frac{1}{\hat{x}}\right)\left(1-\frac{1}{\hat{z}}\right)
+\frac{m_c^2}{\hat{z}^2Q^2}\right)
\nonumber\\
&&
={2q_T\over Q^2}\int{dx\over x^2}
\left[
{ \xhat\hat{\sigma}_k^U(Q,q_T,\xhat,\zhat)\over 1-\zhat}
\left( x{dO(x,x)\over dx}-2O(x,x)\right)\right. 
+\left\{
{Q^2\over \zhat} {\partial \hat{\sigma}_k^U(Q,q_T,\xhat,\zhat) \over \partial q_T^2}
\right.
\nonumber\\
&&\;\;\;\left.\left.
- {\xhat^2 \over 1-\zhat} {\partial  \hat{\sigma}_k^U(Q,q_T,\xhat,\zhat) \over \partial \xhat}
\right\}
O(x,x)
\right]
\delta\left(
\frac{q_T^2}{Q^2}-\left(1-\frac{1}{\hat{x}}\right)\left(1-\frac{1}{\hat{z}}\right)
+\frac{m_c^2}{\hat{z}^2Q^2}\right),
\label{derivqt}
\eeq
where the derivative, $dO(x,x)/dx$, arises as a result of the integration by parts.
Similar results are obtained for the contribution associated with the correlation function
$N(x,x)$, as well as for the corresponding contributions from (\ref{XsecXO}) to 
be combined with $O(x,0)$ and $N(x,0)$.  
This indicates that all four nonperturbative functions of $x$,
$O(x,x)$, $N(x,x)$, $O(x,0)$ and $N(x,0)$, contribute both in the derivative and nonderivative forms
to the twist-3 SSA,
and the partonic hard cross sections convoluted with them are entirely determined from the hard scattering parts
for the $\gamma^* g\to c\bar{c}$ scattering which corresponds to the $2 \to 2$ process at the twist-2 level.

Substituting (\ref{LWS})
into (\ref{diffsigma}) and using (\ref{XsecXX}), (\ref{XsecXO})-(\ref{derivqt}),
we obtain the leading-order QCD formula for the single-spin-dependent cross section $\Delta \sigma$
in the SIDIS, $ep^\uparrow\to eDX$, 
generated from the twist-3 three-gluon correlation functions $O(x_1,x_2)$ and $N(x_1,x_2)$
of (\ref{3gluonO}) and (\ref{3gluonN}) as
\beq
&&
\frac{d^6\Delta\sigma
}{dx_{bj}dQ^2dz_fdq_T^2d\phi d\chi}\nonumber\\
&&
=\frac{\alpha_{em}^2\alpha_s e_c^2 M_N}{16\pi^2 
 x_{bj}^2S_{ep}^2Q^2}\left(\frac{-\pi}{2}\right) 
\sum_{k=1,\cdots, 4,8,9}
\int\frac{dx}{x^2}\int\frac{dz}{z} \delta\left(
\frac{q_T^2}{Q^2}-\left(1-\frac{1}{\hat{x}}\right)\left(1-\frac{1}{\hat{z}}\right)
+\frac{m_c^2}{\hat{z}^2Q^2}\right)D(z)\nonumber\\
&&\times\left(
\sin(\Phi_S-\chi) {\cal A}_k{2q_T\over Q^2}\left[ {\xhat\hat{\sigma}_k^U\over 1-\zhat}
\left(x{dO(x,x)\over dx}-2O(x,x)\right) +\left(
{Q^2\over \zhat}{\partial \hat{\sigma}_k^U \over \partial q_T^2}  
- {\xhat^2 \over 1-\zhat}{\partial \hat{\sigma}_k^U \over \partial \xhat}\right)
O(x,x)\right] \right.  \nonumber\\
&&\left.\qquad
+ \cos(\Phi_S-\chi){\partial {\cal A}_k \over \partial \phi} {\hat{\sigma}_k^U\over \zhat q_T} 
O(x,x) +\left( O(x,x)\to N(x,x)\right) \right.
\nonumber\\
&&\left.\qquad+\sin(\Phi_S-\chi) \left[
-2{\cal A}_k{2q_T\over Q^2}\left\{
{\xhat\hat{\sigma}_k^{XX}\over 1-\zhat}
\left(x{dO(x,0)\over dx}-2O(x,0)\right) 
\right.\right.\right.\nonumber\\
&&\left.\left.\left.\qquad\qquad\qquad\qquad\qquad\qquad
+\left(
{Q^2\over \zhat}{\partial \hat{\sigma}_k^{XX} \over \partial q_T^2}  
- {\xhat^2 \over 1-\zhat}{\partial \hat{\sigma}_k^{XX} \over \partial \xhat}\right)
O(x,0)\right\} \right.\right.\nonumber\\
&&\left.\left.\qquad\qquad
+{1\over \zhat q_T}\left\{
{\partial {\cal A}_k\over \partial \phi}\hat{\sigma}_k^{\{XY\}}
+2{\cal A}_k\left( -\hat{\sigma}_k^{XX} + \hat{\sigma}_k^{YY} \right)
\right\}O(x,0)\right]
\right.\nonumber\\
&&\left.\qquad
+\cos(\Phi_S-\chi) \left[
{\cal A}_k{2q_T\over Q^2}\left\{ {\xhat\hat{\sigma}_k^{\{XY\}}\over 1-\zhat}
\left(x{dO(x,0)\over dx}-2O(x,0)\right) 
\right.\right.\right.\nonumber\\
&&\left.\left.\left.\qquad\qquad\qquad\qquad\qquad\quad
+\left(
{Q^2\over \zhat}{\partial \hat{\sigma}_k^{\{XY\}} \over \partial q_T^2}  
- {\xhat^2 \over 1-\zhat}{\partial \hat{\sigma}_k^{\{XY\}} \over \partial \xhat}\right)
O(x,0)\right\} \right.\right.  \nonumber\\
&&\left.\left.\qquad\qquad
+{1\over \zhat q_T}\left(
-2{\partial {\cal A}_k\over \partial \phi}\hat{\sigma}_k^{YY}
+2{\cal A}_k  \hat{\sigma}_k^{\{XY\}}
\right)O(x,0)\right] +\left( O(x,0)\to - N(x,0) \right)
\right),
\label{finalmaster}
\eeq
where $\alpha_s=g^2/(4\pi)$ is the strong coupling constant,
${\cal A}_k\equiv {\cal A}_k(\phi-\chi)$, 
and $\hat{\sigma}_k^j$ ($j=U, XX, YY, \{XY\}$) are defined in (\ref{Hsigma}).
It is worth noting the relations
$\hat{\sigma}^{j}_{8,9}=0$ for $j=XX,\,YY$ and 
$\hat{\sigma}^{j}_{1,2,3,4}=0$ for $j=\{XY\}$, since even number of $Y^\mu$'s
has to be involved in the contraction to give nonzero $\hat{\sigma}^{j}_k$. 
We also remind the relations
${\partial {\cal A}_3 \over \partial \phi}=-{\cal A}_8$,
${\partial {\cal A}_4 \over \partial \phi}=-2{\cal A}_9$
${\partial {\cal A}_8 \over \partial \phi}={\cal A}_3$ and 
${\partial {\cal A}_9 \over \partial \phi}=2{\cal A}_4$
in (\ref{finalmaster}).
It is straightforward to derive the explicit formulae for $\hat{\sigma}_k^j$ 
by calculating the diagrams for the $2\to 2$ processes in Fig.~3:
For example, $\hat{\sigma}^{U}_k$  ($=-\hat{\sigma}^{XX}_k-\hat{\sigma}^{YY}_k$)
are given in Eq.~(81) of \cite{BKTY10}, where
$\hat{\sigma}^{U}_k$ are nonzero for $k=1,\cdots,4$, while $\hat{\sigma}^{U}_{8,9}=0$. 
Those determine
the gluon contribution to the twist-2 unpolarized cross section $\sigma^{\rm unpol}$
for $ep\to eDX$, as
\begin{eqnarray}
&&\hspace{-0.5cm}\frac{d^5\sigma^{\rm unpol}}{dx_{bj}dQ^2dz_fdq_T^2d\phi}
=\frac{\alpha_{em}^2\alpha_se_c^2}{8\pi
 x_{bj}^2S_{ep}^2Q^2}\frac{1}{4}
\sum_{k=1}^4{\cal A}_k(\phi)\int
\frac{dx}{x}
\int
\frac{dz}{z}\sum_{a=c,\bar{c}}D_a(z)G(x) \hat{\sigma}^{U}_k\nonumber\\
&&\qquad\qquad\qquad
\qquad\qquad\qquad\times\delta\left(\frac{q_T^2}{Q^2}-
\left(1-\frac{1}{\hat{x}}\right)\left(1-\frac{1}{\hat{z}}\right)
+\frac{m_c^2}{\hat{z}^2Q^2}\right),
\label{unpolresult}
\end{eqnarray}
which can be obtained immediately from (\ref{unpol}), (\ref{gxgx}) using (\ref{cav}) and (\ref{Hdef}),
integrating over $\chi$
for the fixed $\phi' \equiv \phi-\chi$, and perfoming the formal replacement $\phi' \rightarrow \phi$
(i.e., $\phi$ in (\ref{unpolresult}) is understood to be this $\phi'$).
We do not show the rather lengthy formulae
for the other hard cross sections in (\ref{Hsigma}).

Note that (\ref{finalmaster}) shows the result for the $c$-quark fragmentation 
channel according to the diagrams in Fig.~3.
The contribution for the $\bar{c}$-quark fragmentation channel, due to the diagrams
in Fig.~3 with the direction of the quark lines reversed,
can be calculated similarly as above,
and yields the formula (\ref{finalmaster})
with the $c$-quark fragmentation function $D(z)$ replaced by the $\bar{c}$-quark fragmentation function 
and also with the replacements $O(x,x) \rightarrow -O(x,x)$ and $O(x,0) \rightarrow -O(x,0)$,
reflecting the fact that (\ref{3gluonO}) is associated with the $C$-odd combination of the gluon operators.
Combining this result with (\ref{finalmaster}),
we obtain the total result,
which can be recast into the following form: 
\beq
&&
\hspace{-0.3cm}\frac{d^6\Delta\sigma
}{dx_{bj}dQ^2dz_fdq_T^2d\phi d\chi}\nonumber\\
&&
=\frac{\alpha_{em}^2\alpha_se_c^2 M_N}{16\pi^2
 x_{bj}^2S_{ep}^2Q^2}\left(\frac{-\pi}{2}\right) 
\sum_{k=1,\cdots, 4,8,9}
{\cal
 A}_k{\cal S}_k
\int
\frac{dx}{x}\int
\frac{dz}{z} \delta\left(
\frac{q_T^2}{Q^2}-\left(1-\frac{1}{\hat{x}}\right)\left(1-\frac{1}{\hat{z}}\right)
+\frac{m_c^2}{\hat{z}^2Q^2}\right)\nonumber\\
&&\qquad
\times \sum_{a=c,\bar{c}}
D_a(z) \left[\delta_a\left\{
\left(\frac{d}{dx}O(x,x)-\frac{2O(x,x)}{x}\right)\Delta\hat{\sigma}^{1}_k
+\left(\frac{d}{dx}O(x,0)-\frac{2O(x,0)}{x}\right)\Delta\hat{\sigma}_k^2
\right.\right.\nonumber\\
&&\qquad\qquad\qquad\qquad\left.\left.
+\frac{O(x,x)}{x}\Delta\hat{\sigma}^{3}_k
+\frac{O(x,0)}{x}\Delta\hat{\sigma}^{4}_k
\right\}\right. \nonumber\\
&&\qquad\qquad\left.
+\left\{
\left(\frac{d}{dx}N(x,x)-\frac{2N(x,x)}{x}\right)\Delta\hat{\sigma}^{1}_k
-\left(\frac{d}{dx}N(x,0)-\frac{2N(x,0)}{x}\right)\Delta\hat{\sigma}^{2}_k\right.\right.\nonumber\\
&&\left.\left.\qquad\qquad\qquad\qquad+\frac{N(x,x)}{x}\Delta\hat{\sigma}^{3}_k
-\frac{N(x,0)}{x}\Delta\hat{\sigma}^{4}_k
\right\}
\right],
\label{3gluonresult}
\eeq
where ${\cal A}_k\equiv {\cal A}_k(\phi-\chi)$, and 
${\cal S}_k$ is defined as
${\cal S}_k=\sin(\Phi_S-\chi)$ for $k=1,2,3,4$
and ${\cal S}_k=\cos(\Phi_S-\chi)$ for $k=8,9$.
The quark-flavor index $a$ can, in principle, be $c$ and $\bar{c}$, with $\delta_c=1$ and  
$\delta_{\bar{c}}=-1$, so that the cross section for 
the $\bar{D}$-meson production $ep^\uparrow\to e\bar{D}X$
can be obtained by 
a simple replacement of the fragmentation function
to that for the $\bar{D}$ meson, $D_a (z) \rightarrow \bar{D}_a (z)$.   
Comparing (\ref{3gluonresult}) with (\ref{finalmaster}), we find 
\beq
\Delta\hat{\sigma}_k^1={2q_T\xhat\over Q^2(1-\zhat)} \hat{\sigma}^U_k,
\eeq
and, similarly, 
the partonic hard cross sections
$\Delta\hat{\sigma}_k^j$ ($j=2,3,4$, $k=1,\cdots,4,8,9$)
are completely expressed by $\hat{\sigma}_k^j$ ($j=U, XX, YY, \{XY\}$) in (\ref{Hsigma}).
Substitution of the above-mentioned explicit forms of $\hat{\sigma}_k^j$
into these relations
yields the formulae for $\Delta\hat{\sigma}_k^j$;
these formulae, of course, 
agree with Eqs.~(71)-(74) in \,\cite{BKTY10},
which were obtained by the direct calculation 
of the three-gluon diagrams in Fig.~2 in
our previous study.

\subsection{Toward extension to higher orders}

As we have demonstrated in the last section,
our master formula 
allows us to derive
the explicit form of the twist-3 SSA for 
$ep^\uparrow \to eDX$ in the leading-order QCD,
taking into account the whole contribution induced by the three-gluon correlation inside the
transversely-polarized nucleon.
Thus, extension of our master formula beyond the leading-order QCD
is interesting in that it would provide a powerful framework that allows us
to calculate the higher-order corrections to the SSA,
which are unknown at present.

To derive the master formula, our starting point was (\ref{wfinal}),
which gives the whole twist-3 cross section  
as the SGP contribution at $x_1=x_2$.   
The derivation of (\ref{wfinal})
presented in \cite{BKTY10}
relies on particular properties 
satisfied by the corresponding partonic scattering amplitudes with an on-shell internal line
for which the propagator is replaced by its pole contribution.   
In particular, those properties include
Ward identities for the relevant partonic amplitudes at the leading order.
We expect that the similar Ward identities hold even after inclusion of the higher-order corrections
employing a suitable gauge choice like the background-field gauge,
and that (\ref{wfinal}) is useful beyond the leading order.
Once the cross section is expressed as in (\ref{wfinal}), it is easy to see that 
the formulae (\ref{master1}), (\ref{master2}) hold
for the SGP contributions:   
We note that the essential ingredient leading to these relations is the 
simple correspondence between the 
hard scattering part (\ref{Stwist3}) relevant to the SGP contributions and the hard part (\ref{Stwist2})
for the $\gamma^*g\to c\bar{c}$. 
As shown in (\ref{Smaster}) and (\ref{master2}), 
for the 
contributions that can be expressed as (\ref{Sleft})-(\ref{Stwist3}) 
using the $\gamma gcc$ vertices ${\cal F}_\alpha^a$
and $\bar{\cal F}_\beta^b$,
the derivative with respect
to $k_2^\gamma$ in (\ref{wfinal}) can be ``transformed'' into
the derivative with respect to $p_c^\gamma$ in the master formula,
without modifying ${\cal F}_\alpha^a$
and $\bar{\cal F}_\beta^b$.
Similarly,
for the higher-order 
contributions that can be expressed generically as (\ref{Sleft})-(\ref{Stwist3}) 
using the $\gamma gcc$ vertices ${\cal F}_\alpha^a$
and $\bar{\cal F}_\beta^b$
with the corresponding corrections included, 
the result (\ref{master2}) holds 
without additional modification to ${\cal F}_\alpha^a$
and $\bar{\cal F}_\beta^b$.

When including higher-order corrections, the SGP contributions 
may also occur from the diagrams that contain, in general, a
number of final unobserved partons,
as shown in Fig.~4.  
In that case, the vertex ${\cal F}_\alpha^a$ in Fig.~4 has
a number of external legs together with the loop correction inside. 
Still, we note that
only the diagrams, which have the extra gluon-line attached to the 
final parton fragmenting into the $D$-meson,
eventually contribute to the twist-3 cross section\,\cite{QS92}.  
Because of this particular structure in the relevant SGP contributions,
we can show that the relation (\ref{master2}) and thus our master formula hold for the 
twist-3 single-spin-dependent cross section for 
$ep^\uparrow \to eDX$, 
even after the inclusion of the
higher-order corrections. The details will be discussed elsewhere.

\begin{figure}[t!]
\begin{center}
\epsfig{figure=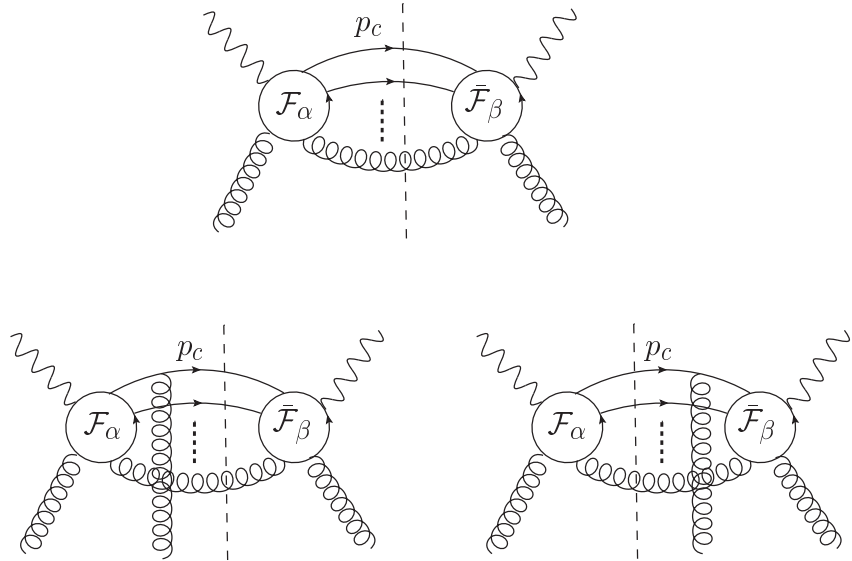,width=0.7\textwidth}
\end{center}
\caption{Generic diagrams for the higher-order corrections to the 
twist-2 unpolarized cross section
(upper diagram) and the twist-3 single-spin-dependent cross section (lower diagrams).  
\label{fig4}
}
\end{figure}

\section{Summary}
In this paper, we 
have derived the master formula for the contribution of the 
three-gluon correlation functions
to the twist-3 single-spin-dependent cross section for the 
$D$-meson production in SIDIS, $ep^\uparrow\to eDX$.  
This formula connects the twist-3 effects due to the interference arising in the partonic 
hard cross section to 
the Born cross sections for the $\gamma^*g\to c\bar{c}$ scattering at the twist-2 level.   
In particular, the hard cross sections for the three-gluon correlation functions $\{O(x,x),N(x,x)\}$
are completely
determined by the hard cross sections associated with the 
gluon density distribution in the twist-2 unpolarized cross section for $ep\to eDX$.   
In the similar master formula derived for the contribution of the twist-3 quark-gluon correlation functions, 
only the SGP component of the cross section was connected to the twist-2 unpolarized cross section.
For the present case with the three-gluon correlation functions, all contributions to the corresponding cross section appear as the SGP contribution, and thus the master formula derived here is for 
the total twist-3 cross section. 
The formula derived here can be easily extended to the three-gluon
contribution to $p^\uparrow p\to DX$\,\cite{Koike:2011mb}.  
The derivation of the formula is
based on the general structure 
and properties of the relevant twist-3 hard scattering part, which are expected to hold even after 
including the 
higher order corrections.

\section*{Acknowledgments}
The work of S. Y. is supported by the Grand-in-Aid for Scientific Research
(No. 22.6032) from the Japan Society of Promotion of Science.
The work of K.~T. 
is supported in part by
the Grant-in-Aid for Scientific Research on Priority Areas
No.~22011012.


\end{document}